# Engineering the directionality of hot carrier tunneling in plasmonic tunneling structures

Mahdiyeh Abbasi[1], Shusen Liao[2,3], Yunxuan Zhu[3], and Douglas Natelson[1,3,4,*]

[1]Department of Electrical and Computer Engineering, Rice University, Houston, TX 77005, USA

[2]Applied Physics Graduate Program, Smalley-Curl Institute, Rice University, Houston, TX 77005, USA

[3]Department of Physics and Astronomy Engineering, Rice University, Houston, TX 77005, USA

[4]Department of Materials Science and Nanoengineering, Rice University, Houston, TX 77005, USA

[*]Corresponding author: Douglas Natelson (natelson@rice.edu.)

## Abstract

Tunneling metal-insulator-metal (MIM) junctions can exhibit an open-circuit photovoltage (OCPV) response under illumination that may be useful for photodetection. One mechanism for photovoltage generation is hot carrier tunneling, in which photoexcited carriers generate a net photocurrent that must be balanced by a drift current in the open-circuit configuration. We present experiments in electromigrated planar MIM structures, designed with asymmetric plasmonic properties using Au and Pt electrodes. Decay of optically excited local plasmonic modes preferentially creates hot carriers on the Au side of the junction, leading to a clear preferred directionality of the hot electron photocurrent and hence a preferred polarity of the resulting OCPV. In contrast, in an ensemble of symmetric devices constructed from only one Au, polarity of the OCPV has no preferred direction.

Since the photoelectric effect was first observed by Heinrich Hertz in 1887[1], photoemission and related phenomena have been utilized in photodetection applications. Metal-based photoemissive devices can potentially work over a wide spectrum, provided the photoexcited carriers can overcome barriers to emission and be captured. Assuming a gapless density of states[2], the hot electrons generated in metals by photons can range in energy from the Fermi level to a maximum of the photon energy above the Fermi level. Tunnel junctions provide a means of detecting photoexcited hot carriers in the form of a hot carrier tunneling photocurrent[3]. Controlling hot carrier generation and tunneling opens opportunities for novel photodetectors[4] and energy harvesting[5].

Plasmon response provides a means of engineering the generation of hot carriers through design of resonant structures[6–8]. Plasmon modes can decay nonradiatively through Landau damping[9–12], in which each plasmon quantum produces one electron-hole pair. When the energy of these carriers deviates from the thermal equilibrium distribution, these are called 'hot carriers'[9]. The characteristic timescale for electron-electron thermalization is tens of femtoseconds, and picoseconds for inelastic relaxation to the lattice. For an intraband transition, electrons are generally excited from the conduction band to higher energy states in the same band. The excitation can also take place between the conduction band and other bands (e.g., $d$ bands) in a process known as interband excitation. The ratio of electron to hole generation depends on the electronic structure and photon energy. Hot carriers can be used in local heating[13,14], photochemistry[15,16], and photo-desorption where the energy of hot electrons can be used to photo-desorb small molecules form the surface[17]. Hot carriers can also be photoemitted from a metal over either a Schottky[18,19] or oxide tunnel-barrier[20,21].

In this work, we examine hot carrier tunneling in a planar MIM junction between dissimilar metals with different plasmonic properties, through the measurement of the open-circuit photovoltage (OCPV) that results under illumination. If illumination causes a flow of electrons from one side of the device to the other via an optically driven hot carrier current, in the open-circuit configuration an OCPV will build up so that the drift current from the potential difference balances the optically driven carrier motion so that there is zero net current in the system in the steady state. We also examine the systematic OCPV as a function of tunneling conductance, a proxy for closest interelectrode separation. In these devices, the polarity of the OCPV caused by tunneling of these hot carriers is consistent with hot electron tunneling direction based on hot carrier generation from plasmon decay. By designing junctions with a Au electrode that is plasmonically resonant with the excitation wavelength (1060 nm or 785 nm) and the other electrode made from non-resonant Pt, the sign of the OCPV is consistent with photoexcited electrons tunneling from the Au to the Pt. Conversely, when both electrodes are Au and have similar plasmonic response, there is no preference in the sign of the OCPV. In previous studies where the similar metals were used in planar MIM junctions, the polarity of the open circuit photovoltage was not controllable under either direct or indirect optical excitation[22,23]. The MIM tunnel junctions presented here have higher stability than previous structures even at ambient conditions[22,23]. Tunneling structures in principle can be used to make photodetectors that are faster and have higher responsivity comparing to photothermoelectric (PTE) effect-based photodetectors that require the time scale of electron-phonon and phonon-phonon interactions to operate in steady state[24].

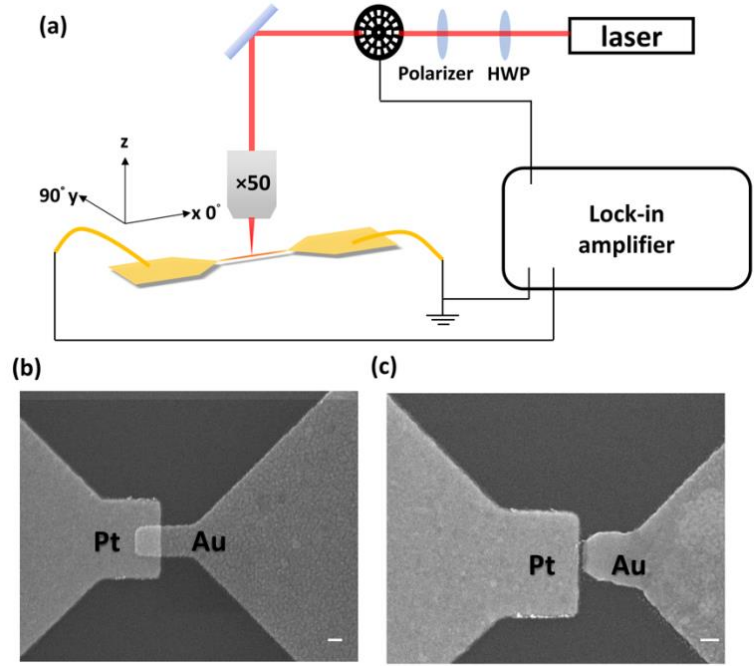

**FIG. 1.** (a) Scheme of the OCPV measurement setup. The laser intensity and polarization are controlled by a half wave plate (HWP) and a polarizer. The laser is modulated by an optical chopper which connects the lock-in amplifier for frequency reference. The beam is focused on the sample by a 50× objective. The OCPV is measured by the lock-in amplifier. (b) The SEM image of the Au-Pt nanowire before electromigration. The Pt and Au nanowires are ~500 and 230 nm wide respectively. (c) The SEM image of the Au-Pt nanowire after electromigration. A sub-nanometer gap is formed at the interface of Au and Pt. The scale bars in (b) and (c) are 100 nm.

All devices are fabricated on Si chips with 2 μm thick $SiO_2$ on the surface. For Au-Pt devices, the Pt electrode side is defined by EBL and 18 nm Pt is sputtered, followed by liftoff. Then a Au nanowire with large contact is defined, aligned to the Pt side, by e-beam lithography (EBL) and liftoff processing. After developing, Au of 18 nm thickness is evaporated with no adhesion layer, to maximize plasmonic response[12]. The nanowires are connected to extended 50 nm thick Au contact pads with a 5 nm thick Ti adhesion layer fabricated by shadow mask e-beam evaporation. Wire bonding is used to connect the large contact pads using gold wires and the gold electrode is connected to the ground (Fig. 1(a)). A scanning electron microscopy (SEM) image of a pre-migration Au-Pt nanowire junction is shown in Fig. 1(b). Subsequently the nanowires are electromigrated at room temperature to form the MIM junction[25]. An increasing voltage bias is applied to the nanowire and the current is monitored as feedback simultaneously. The bias increases by step until the current drops, indicating an increase in resistance. Then the bias is set to 0 to start a new cycle. The cycles are repeated until the conductance of the device is smaller than conductance quantum $G_0 \equiv 2e^2/h$. The SEM image of a junction after electromigration is shown in Fig. 1(c). A sub-nanometer gap is formed at the interface of Au and Pt. Note that the Pt has a higher melting point than Au and is more refractory, so that during electromigration, only the gold nanowire is electromigrated and Pt nanowire remains apparently unchanged. Based on fitting to the Simmons model[26], the nanogap formed in devices presented here is ~1 nm.

The OCPV measurement setup is illustrated in Fig. 1(a). We use a 1060 nm continue wave (CW) laser as the optical source for measurements under ambient conditions, and for measurements under vacuum, we use a 785 nm CW laser source. A combination of half wave plate and polarizer is used to control the intensity and polarization of the laser and the laser is modulated by an optical chopper. The chopper modulation period is much longer than the thermal timescales of the structures[27], so all the data is measured in the steady state. The laser is focused by a Zeiss Epiplan-Neofluar 50x objective with a NA of 0.55 and hits the devices at the focal plane, where laser spot diameters for the 1060 and 785 nm CW lasers are 3.5 and 2.7 μm, respectively, as measured via knife-edge. The open circuit voltage of the device is amplified through the SRS 560 low noise voltage amplifier and then measured via the lock-in amplifier. Two Thorlabs DRV001 stepper motors control the 2D movement of the sample stage in the plane perpendicular to laser direction to change the relative position between laser spot and the device. The OCPV maps (Fig.

2) are obtained by scanning the laser spot over an area with user-defined pixel size. The OCPV $V_{OC}$ is measured at each pixel.

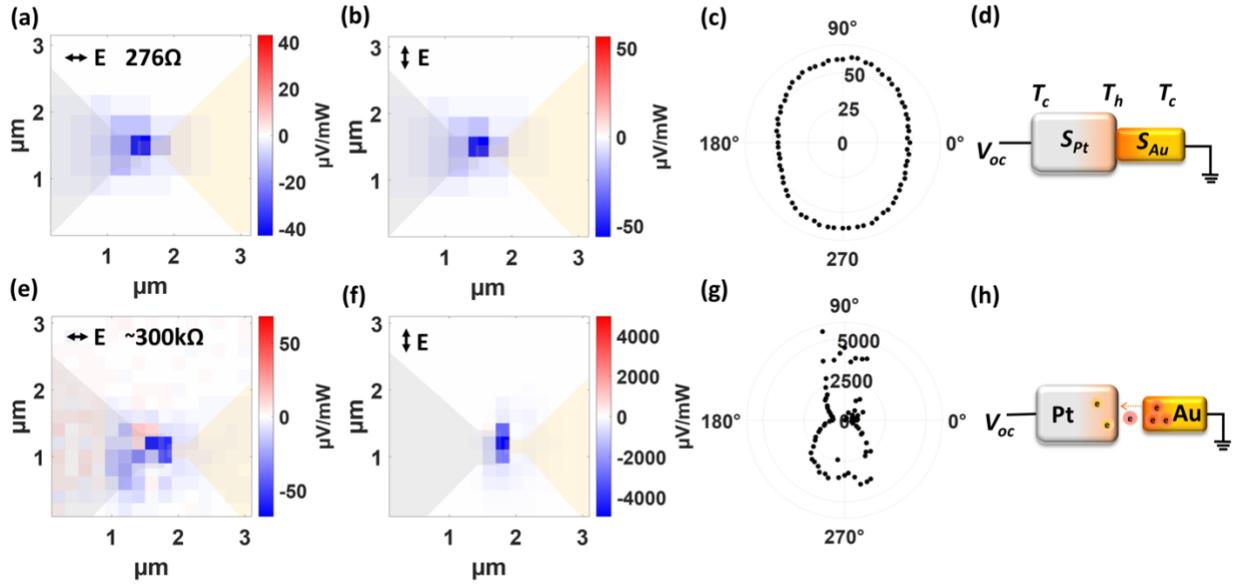

**FIG. 2. Polarization dependent OCPV map and polar plot.** (a) $V_{OC}$ map for 0 degree polarization before electromigration. (b) $V_{OC}$ map for 90 degree polarization before electromigration. The maps contain 10×10 pixels and the size of each pixel is 0.3 µm. The gray and yellow part indicate the positions of Pt and Au nanowires. (c) polar plot of $V_{OC}$ before electromigration. The laser is positioned at the pixel with the largest signal. (d) Scheme of mechanism of OCPV before electromigration. PTE effect is the dominant mechanism. (e) $V_{OC}$ map for 0 degree polarization after electromigration. (f) $V_{OC}$ map for 90 degree polarization after electromigration. The maps contain 15×15 pixels and the size of each pixel is 0.2 µm. (g) polar plot of $V_{OC}$ after electromigration. (h) Scheme of mechanism of OCPV after electromigration. Hot electron tunneling is the dominant mechanism.

The photovoltage measurements of the same device are performed both before and after the electromigration for 0 and 90 degree (parallel and transverse to the nanowire) electric field polarization. The resistance of the device is 276 Ω prior to electromigration and about 300 kΩ after the electromigration. The results are shown in Fig. 2. The PTE signal and the post-migration OCPV, ascribe to hot electron tunneling, are observed to change linearly with applied incident power[22,28], and the measured $V_{OC}$ is normalized to the incident optical power. $V_{OC}$ is negative for both before and after electromigration no matter the polarization is 0 or 90 degree, as shown in Fig. 2(a), 2(b), 2(e) and 2(f). The mechanism for negative $V_{OC}$ is different before and after electromigration. Before electromigration, the PTE effect is the dominant mechanism: the negative photovoltage in Au-Pt devices is due to the Seebeck effect and the thermocouple formed by the Au-Pt junction, which can be descripted by $V_{OC} = (S_{Pt} - S_{Au}) \times (T_{hot} - T_{cold})$. Here, $T_{hot}$ is the temperature at the laser spot position and $T_{cold}$ is the temperature of ambient condition. The laser induces a temperature gradient and the Seebeck coefficient $S_{Au}$ is larger than $S_{Pt}$, which causes $V_{OC}$ to be negative, as illustrated in Fig. 2(d). After electromigration to form a complete nanogap, the $V_{OC}$ increases by about 3 order of magnitude, from µV to mV. Based on the PTE physics, the dramatic increase $V_{OC}$ would imply a 2~3 order of magnitude change in $S$ or the local temperature gradient, which is not physically reasonable[22]. Instead, in the nanogap limit the photovoltage is dominated by hot carrier tunneling, as described in previous work[22]. The incident light excites local surface plasmon (LSP) at the MIM junction. The LSP can be damped non-radiatively through Landau damping and create hot carriers on femtosecond time scale[9–11]. In our devices, more hot carriers are generated at the Au side, which is more plasmonically active, and this causes an asymmetric electron distribution across the junction so the net tunneling current is predominantly from Au to the Pt electrode. The calculated plasmonic absorption of the Pt nanowire is shown in the Supplementary Material. The built-in electric field caused by the work function difference in the asymmetric MIM structure should tend to favor electrons tunneling from Pt to the Au electrode[28]. However, the directional effect by the built-in electric field in the insulator in MIM structures is dominant when the mean free path of electron-phonon scattering in the insulator is smaller than the insulator thickness[28], which is not true for our devices where the gap is in vacuum or air and the gap size is smaller than 1 nm. Moreover, even though the density of states (DOS) of Pt is larger than the DOS of Au near the Fermi level[29], it has been shown previously that the DOS does not force the hot carrier tunneling direction[28]. To counteract the dominant hot carrier tunneling from Au to Pt, the OCPV that the system builds should be negative, as illustrated in Fig. 2(h).

After the map scans, the laser spot is moved to the pixel where the signal is the largest, then the polarization plots are measured, as shown in Fig. 2(c) and 2(g) for before and after electromigration, respectively. Before electromigration, $V_{OC}$ that originates from PTE signal is slightly stronger for transverse (90 degree) light polarization. The transverse polarized light excites the dipolar transverse plasmon mode of the nanowire, which causes larger absorption and a stronger heating effect. This makes the temperature difference $T_{hot} - T_{cold}$ larger, so the $V_{OC}$ for perpendicular polarization is larger. After electromigration, due to symmetry breaking of the gap geometry, the higher order plasmon modes can couple to the transverse dipolar mode through hybridization[30] and excited via transverse polarized light. This results in highly localized, intensely enhanced electric fields,[8] localized generation of hot carriers, and hot carrier tunneling, leading to an ~80 times larger $V_{OC}$ signal than before migration. Note that the gap becomes unstable under constant illumination, so the polar plot after electromigration is not perfectly symmetric. Lower optical power is used on tunnel junctions to improve measurement stability. The much larger $V_{OC}$ after the nanogap formation is consistent with previous reports[22] and is indicative that the OCPV mechanism has changed. The resistance of the nanowire after electromigration is much larger, so a consequently larger OCPV is necessary to balance the optically driven hot carrier tunneling current.

To confirm the plasmonic origin of the OCPV after electromigration, another kind of Au-Pt device is fabricated. The width of the Pt side is fixed at ~500 nm, and width of the Au nanowire is 160 nm. The simulated Au nanowire absorption as a function of Au nanowire width under 1060 nm laser excitation is shown in Fig. 3(a). The simulation results are obtained by the Finite Element Method (FEM) using COMSOL Multiphysics with a simplified 2D planewave illumination model. (Plasmon resonances computed using a full 3D model of a gaussian beam spot are qualitatively identical.) Devices with 230 nm width Au wire (red triangles) have larger absorption and are closer to the plasmon peak than devices with 160 nm width Au wire (blue circles). As a result, for the devices with wider Au nanowire, there should be more hot electrons generated at the Au side and the $V_{OC}$ is expected to be larger, all other things being equal. An ensemble of devices (19 devices with 230nm width Au wire, 21 devices with 160 nm width Au wire) are measured. $V_{OC}$ is about 5 times larger for 230 nm devices than 160 nm devices, as shown in Fig. 3(b). The influence of the increased geometrical cross section for wider devices is not responsible for the much larger OCPV signals. See Supplementary Material for more details. This supports the important role of plasmons in the photovoltaic response of these MIM junctions.

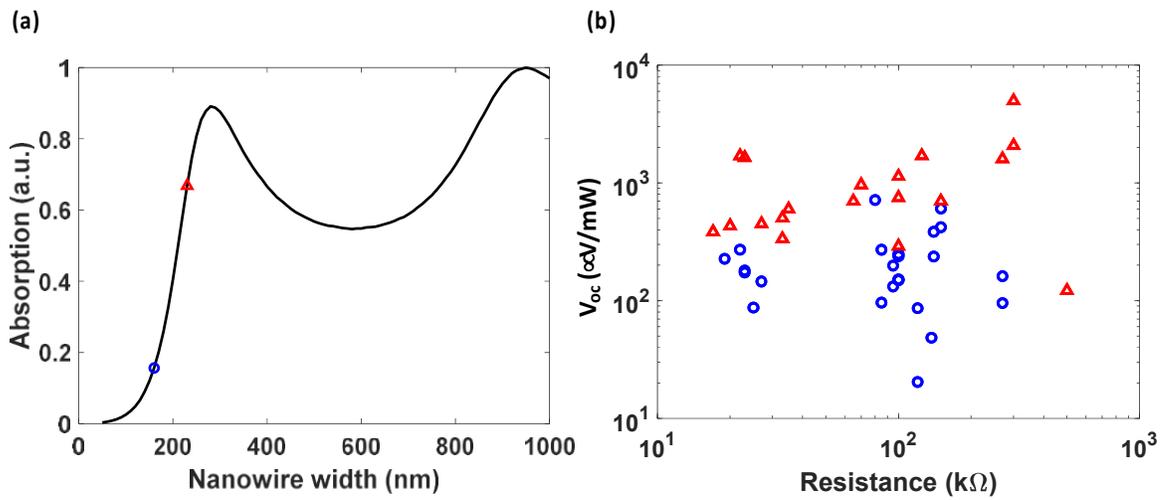

**FIG. 3.** (a) Simulated absorption of the pure Au nanowires with different width under 1060 nm laser illumination. The red triangle and blue circle mark the devices with 230 and 160 nm width Au wire. (b) $V_{OC}$ for an ensemble of Au-Pt devices. The wider Au devices have systematically larger OCPV values.

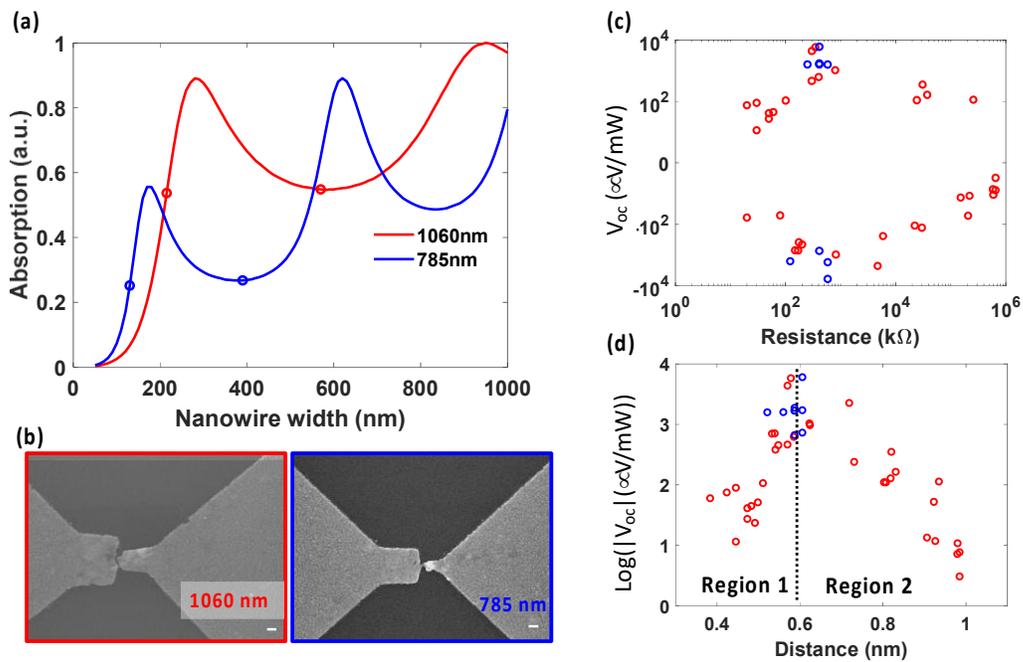

**FIG. 4.** (a) Simulated absorption of the pure Au nanowires with different width under 1060 nm laser (red curve) and 785 nm laser (blue curve) illumination. (b) The SEM images of devices after electromigration

designed respectively for the 1060 and 785 nm lasers. The width of two sides is marked by the red (1060 nm) and blue circles (785 nm) in (a). The width of the two sides is chosen to make the absorption be roughly the same. The scale bars are 100 nm. (c) $V_{OC}$ for an ensemble of pure Au devices. (d) log of open circuit voltage versus the estimated gap distance. The plot is divided into two regions. The red and blue circles are devices designed for 1060 and 785 nm laser respectively.

The OCPV of pure gold devices is measured for a comparison to further confirm that the directionality of electron tunneling in Au-Pt MIM junctions originates from the different plasmonic properties of Au and Pt. The $V_{OC}$ is measured under two conditions: ambient conditions with 1060 nm laser excitation, and room temperature in vacuum with 785 nm laser excitation. The geometries of pure Au devices are designed to be different for different laser illumination, but in both cases the dimensions are chosen to achieve approximately the same absorption on both sides of the junction. The absorption of Au nanowires with different widths under 1060 nm and 785 nm illumination is simulated in Fig. 4(a). The narrow side has a stronger dipolar plasmon mode and the wider side has a larger volume, leading to approximately the same total absorption expected for each side of the junction. The SEM images of the nanowires after electromigration are shown in Fig. 4(b). $V_{OC}$ of 43 pure Au devices is measured, as shown in Fig. 4(c). The devices measured under 1060 nm and 785 nm laser are marked by red and blue circles, respectively. With the resistance of the devices $R$ varying from around 10 to $10^6$ k$\Omega$, the polarity of $V_{OC}$ in pure gold devices is roughly half positive and half negative, which means from device to device the hot carriers randomly tunnel from thin side to wider side (negative polarity) or the other way (positive polarity) when the plasmon response of the two sides is almost the same. For an individual pure Au device, the polarity of $V_{OC}$ depends on the details of the local plasmon modes generated at the asymmetric junction. The detailed local surface plasmon modes at the nanogap are determined by the geometry of the device, which is formed during electromigration process and does not bias any particular direction of hot carrier generation or tunneling.

We qualitatively consider the nonmonotonic dependence of the magnitude of $V_{OC}$ on the zero-bias tunneling conductance shown in Fig. 4(c). We can define the gap distance $d$ to be the internuclear distance between closest atoms, and this can be estimated by the equation $G = G_0 \exp(-\beta(d - d_0))$, where $G$ is the conductance of the junction; $G_0 \equiv 2e^2/h$ is the conductance quantum; $\beta$ is the attenuation factor and $d_0$ is the lattice constant. For Au, $\beta$ = 18.5 nm$^{-1}$ and $d_0$ = 0.4 nm[31]. A rough estimate of the gap distance $d$ inferred from the measured conductance ranges from ~0.35 nm to 1nm. A plot of log of open circuit voltage versus the estimated gap distance is shown in Fig. 4(d). The plot can be understood in two limits. In region 1, when the gap is getting larger, the resistance across the junction increases. The ohmic expression $V_{OC} = R \times I_p$, links the open-circuit voltage to the tunneling resistance $R = 1/G$ the hot electron tunneling current $I_p$ driven by the optical excitation[22,32]. If the magnitude of the hot electron current is dominated by the rate of plasmon-decay-based generation of carriers, over small changes in gap size $I_p$ can credibly be roughly constant. The increase in $R$ then causes the open circuit voltage to increase as the gap is widened. The $V_{OC}$ continues increasing until $R$ reaches ~300 kΩ to 500 kΩ, which correspond to a rough gap distance of 0.57 nm to 0.6 nm. In region 2 we conjecture that the rate limiting process is no longer the generation of hot carriers but the tunneling probability, the decreasing exponentially with the increasing size of the gap[22,32]. The non-monotonic relationship between resistance and $V_{OC}$ is seen in the aggregate of a population of junctions with a large variation of the $R$ (10 ~ 10$^6$ kΩ). The Au-Pt junctions span a much smaller range of tunneling resistances, so that such a dependence is not apparent in Fig. 3(b). A full, detailed theoretical model of the distance dependence of the hot electron OCPV is beyond the scope of this work.

In this work, we measure the OCPV caused by hot electron tunneling in asymmetric Au-Pt MIM devices. All devices show a consistent polarity, indicating that the hot electrons are generated on the more plasmonically resonant Au side and tunnel to the Pt side. By measuring the OCPV of Au-Pt devices with different widths of Au wire, we confirm that plasmons play an important role in generating hot carriers and tunneling current in the MIM junctions. There is no preferred polarity of the OCPV of pure Au devices with similar plasmonic absorption on the two sides of the junction, consistent with the plasmonic properties of the two sides of planar MIM junctions determining the directionality of hot electron tunneling and OCPV. The results demonstrate that plasmonic properties can be used to manipulate directionality of hot carrier tunneling in planar MIM junctions, relevant to potential future applications of integrated, on-chip, plasmonic optoelectronic devices.

*Supplemental material*

The supplemental material includes discussion and plots concerning the role of device geometry in the calculated absorption of the Au and Pt structures.

## ACKNOWLEDGMENT

D.N., S.L., and Y. Z. acknowledge Robert A. Welch Foundation Award C-1636. D.N. and M.A. acknowledge NSF award ECCS-1704625.

## AUTHOR INFORMATION

**Corresponding Author**

*Email: natelson@rice.edu

**Author Contributions**

D.N. and M.A. designed the experiment. M.A. fabricated the devices. MA and Y.Z. conducted the experiment. M.A. and S.L. modelled the data. D.N., M.A. and S.L. wrote the manuscript and have given approval to the final version of the manuscript.

**Notes**

The authors declare no competing financial interest.

**Funding Sources**

Robert A. Welch Foundation Award C-1636.

NSF ECCS-1704625

**Data Availability**

The data that support these findings are available on Zenodo (https://doi.org/10.5281/zenodo.7987316).

# Supplementary Materials

# Engineering the directionality of hot carrier tunneling in plasmonic tunneling structures

Mahdiyeh Abbasi[1], Shusen Liao[2,3], Yunxuan Zhu[3], and Douglas Natelson[1,3,4,*]

[1]Department of Electrical and Computer Engineering, Rice University, Houston, TX 77005

[2]Applied Physics Graduate Program, Smalley-Curl Institute, Rice University, Houston, TX 77005

[3]Department of Physics and Astronomy Engineering, Rice University, Houston, TX 77005

[4]Department of Materials Science and Nanoengineering, Rice University, Houston, TX 77005

[*]Corresponding author: Douglas Natelson (natelson@rice.edu.)

# 1. Geometric cross section effect of Au nanowire for the OCPV

To have a conservative estimation, we assume the plasmon response strength of 160 and 230 nm wide devices to be the same, and the larger physical cross section should give a larger current by the factor $230/160 = 1.77$. However, the OCPV is about 5 times larger for wider devices, so the majority contribution to the much larger OCPV is the stronger plasmon response. We normalize the absorption curve in Fig. 3a by the width of nanowire as the following Fig. S1. The normalized absorption is larger for Au nanowire with 230 nm width, which means that the plasmon response is indeed stronger.

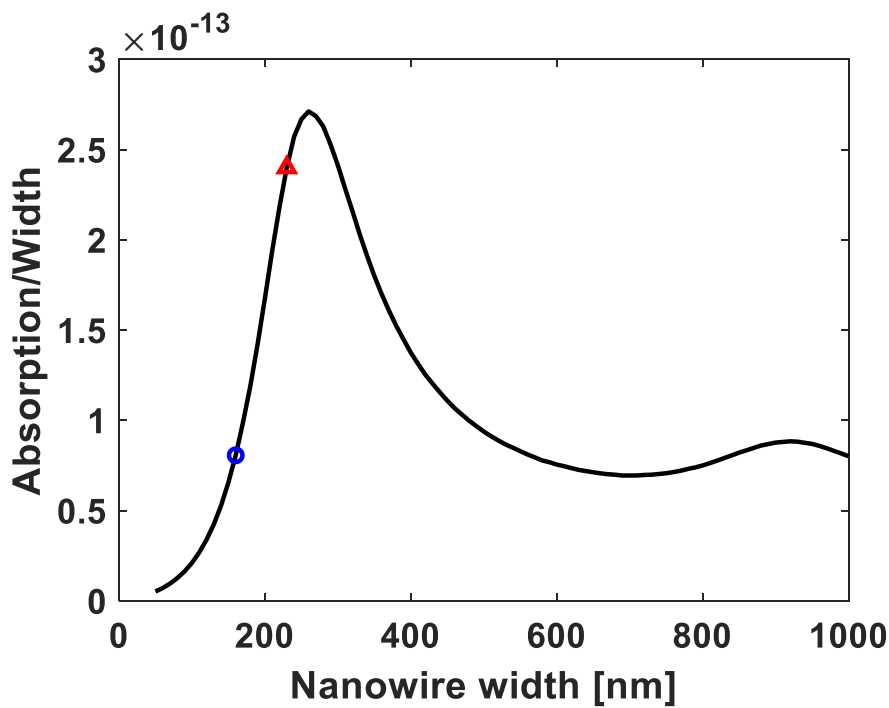

**Fig. S1**. Calculated absorption normalized by nanowire width, to account for purely geometric factors. Even when accounting for the greater width, the 230 nm width nanowire (red triangle) has considerably greater absorption than the 160 nm width nanowire (blue circle), a difference that must come from plasmonic resonant response.

# 2. Plasmonic absorption of Pt nanowire.

The absorption dependence on width for a Pt nanowire under 1060 nm laser illumination is simulated via COMSOL, as shown below in Fig. S2, assuming transverse polarization (looking at a possible transverse plasmon resonance in the Pt, as in the Au nanowire). The 500 nm wide Pt wire also absorbs photon energy a lot, slightly less than the 230 nm wide Au wire (panel a). For 1060 nm laser illumination, the plasmon response strength for Au is stronger for almost all nanowire width values (panel b, blue curve is higher than black curve). From absorption normalized by the width, 230 nm wide Au wire has much stronger plasmon response, and the strength for 500 nm Pt wire and 160 nm Au wire is almost the same. However, we still see the negative value OCPV for 160 nm wide Au devices, which suggests that although the energy absorbed by the Pt and Au sides are almost the same, the efficiency of non-radiative decay of LSPs to hot electrons in the two materials must be different.

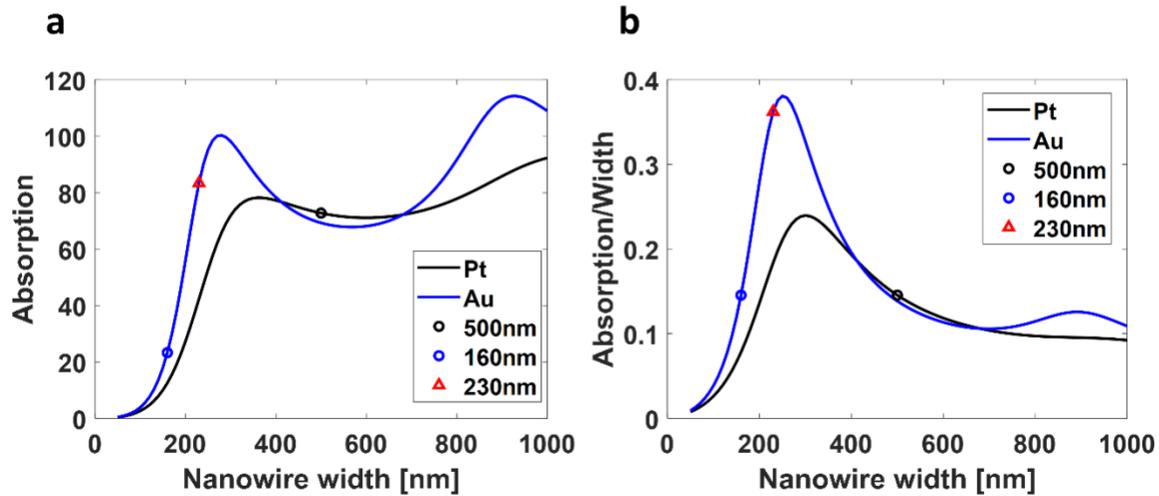

**Fig. S2**. Calculated absorption (for transverse polarized planewave illumination at 1060 nm) comparing Au and Pt wires of indicated widths. While the Pt wire does have a broad plasmonic resonance, the absorption normalized by width does show that plasmonic absorption in the Au case is more prominent. The fact that the OCPV in Au/Pt structures is always consistent with electrons flowing from Au to Pt even in the 160 nm Au width case (when Au absorption should be lower) implies that actual hot carrier production is different in the two materials.